# Particles as probes for complex plasmas in front of biased surfaces

*R. Basner\*, F. Sigeneger\*, D. Loffhagen\*, G. Schubert\*\*, H. Fehske\*\*, H. Kersten\*\*\**

*\* Leibniz-Institute for Plasma Research and Technology (INP) Greifswald*

*\*\* Institute for Physics, University of Greifswald*

*\*\*\* Institute for Experimental and Applied Physics, University of Kiel*

**Abstract**

An interesting aspect in the research of complex (dusty) plasmas is the experimental study of the interaction of micro-particles with the surrounding plasma for diagnostic purposes. Local electric fields can be determined from the behaviour of particles in the plasma, e.g. particles may serve as electrostatic probes. Since in many cases of applications in plasma technology it is of great interest to describe the electric field conditions in front of floating or biased surfaces, the confinement and behaviour of test particles is studied in front of floating walls inserted into a plasma as well as in front of additionally biased surfaces. For the latter case, the behaviour of particles in front of an adaptive electrode, which allows for an efficient confinement and manipulation of the grains, has been experimentally studied in dependence on the discharge parameters and on different bias conditions of the electrode. The effect of the partially biased surface (dc, rf) on the charged micro-particles has been investigated by particle falling experiments. In addition to the experiments we also investigate the particle behaviour numerically by molecular dynamics, in combination with a fluid and particle-in-cell description of the plasma.

# 1. Introduction

Complex (dusty) plasmas which can form plasma or Coulomb crystals [1-3] are at present a topical research subject in plasma physics [4]. Dusty plasmas exhibit a quite heterogeneous composition: neutrals, ions, electrons, and dust particles. This complexity results in complicated interactions at different scales in energy, space, time, and mass. Prominent examples are the formation of structures in particle-containing plasmas under microgravity [5] or the formation of so-called Coulomb balls [6].

Such experimental and theoretical studies initiated the idea to use externally injected small particles, which are negatively charged and affected by several forces in plasma, as micro-probes [7,8]. From the behaviour of the particles in the surrounding plasma local electric fields can be determined ("particles as electrostatic probes") [8,9]. Moreover, energy fluxes towards the particles ("particles as thermal probes") [10,11], or reactive processes on surfaces ("particles as micro-substrates") [12,13] are worth to be studied.

Since the micro-particles can be observed in the plasma sheath easily, they can serve, in particular, as electrostatic probes for the characterisation of the potential surfaces and electric fields in this region [7,9,14]. Usually, the plasma sheath – which is an important zone of energy consumption and, hence, often the essential part of a discharge for application – is difficult to monitor by common plasma diagnostics as Langmuir-probes or optical spectroscopy. Monitoring the position and movement of the particles in dependence on the discharge parameters, information can be obtained on the electric field in front of electrodes and substrate surfaces where other plasma diagnostic methods fail.

If dust microscopic particles are injected into a plasma, they become negatively charged up to the floating potential $V_{fl}$ by electron and ion currents ($j_e$, $j_i$) towards the particles, and can be confined in the discharge. The spatial distribution and movement of the dust particles in a low-temperature plasma is a consequence of several forces acting on the particles [2,7,15].

The charged particles interact with the electric field in front of the electrode or wall, respectively, where the electrostatic force has to be balanced by various other forces in order to confine the dust grains. These forces, which have been discussed extensively by several authors [15-17], are gravitation, neutral and ion drag, thermophoresis and photophoresis. In a variety of process plasmas the electrostatic force, which is proportional to the electric field strength in the sheath, is the dominant force in comparison to the others. Hence, the use of charged micro-particles to obtain additional information on the sheath structure has been successfully demonstrated in front of the powered electrode of capacitively coupled rf-discharges [7,14,18]. In this paper, the behaviour of test particles is described under variation of plasma parameters and different conditions of the surrounding surfaces. Especially, we present studies on the behaviour of charged dust grains in front of grounded or biased but not powered electrodes, where the field is usually stronger. These situations are of special interest for plasma processing of substrate surfaces [19,20] and in plasma chemistry [21].

In view of a theoretical description of the experimental findings, we apply a fluid model as well as particle in cell (PIC) techniques to model the plasma discharge, focussing on the characteristics in the sheath. The micro-particles in the plasma are investigated by means of a molecular dynamics (MD) simulation, using either empirically determined potentials or the PIC results as input. A two-dimensional axisymmetric fluid model comprising the whole reactor volume was used to reveal the local enhancement of the plasma in the vicinity of an rf biased surface. Furthermore, the modification of the electric potential and charge carrier densities was modelled by two- and three-dimensional PIC simulations in the volume near the central pixel of an adaptive electrode.

## 2. Dust particles as electrostatic micro-probes in a plasma sheath

According to the balance of gravitational force ($F_g$), electrostatic force ($F_{el}$), ion drag ($F_i$), neutral drag ($F_n$), thermophoresis, and Coulomb interaction, micro-particles disperse in a relatively small region of the plasma sheath depending on their size and charge. But only some of these forces will play a role in laboratory complex plasmas under certain conditions. Commonly, the electrostatic and gravitational forces are important. Superposition of both forces results in a harmonic potential trap around an equilibrium position [14,15].

The knowledge and the manipulation of the spatial particle distribution is of great interest, for example, for sorting of particles and surface modification of powders [22-24]. However, the interaction of the plasma with injected dust particles is of interest not only regarding their spatial distribution. Vice versa, from the particle behaviour conclusions about the surrounding plasma and sheath properties can be obtained, e.g., field strength and structure.

Since dust grains are small isolated substrates in a plasma environment, they always attain the floating potential. As electrons are much more mobile than ions, the grain surface collects a negative charge, repelling electrons and attracting positive ions until a stationary state is reached. As a result, the net charge $Q = -Ze_0$ of a micron-sized particle can be in the order of a few thousands elementary charges $e_0$. In principle, the charge $Q$ on a micro-particle can be obtained by equating the fluxes of electrons and positive ions towards the particle surface and their recombination [7,25,26]. Several authors have studied both theoretical and experimental aspects of charging the dust grains in capacitively coupled rf discharges, see for example refs. [2,7,27,28].

The charged particles interact with the electric field in front of the electrodes or other surfaces and are often observed as levitated dust clouds forming rings or domes in the boundary regions of the plasma. For several applications in plasma technology as etching or deposition of thin films, modification of powder or composite materials the micro-particles are not confined in front of powered electrodes but in front of surfaces which are additionally biased. Therefore, in the next chapters we discuss the particle behaviour and its use for sheath diagnostics in front of (i) floating surfaces which are inserted into the plasma, (ii) dc-biased surfaces, and (iii) rf-biased surfaces.

## 2.1. Particle behaviour in front of floating surfaces

In dusty plasma experiments fine particles are usually levitating in the horizontal plane above a metal electrode and show a spatial distribution, which depends on the electric field structure above the electrode. Under some conditions, vortices appear and the micro-particles move in the plasma [29,30]. These motions are often generated by surfaces of different potential. Commonly, micro-particles are negatively charged in plasmas. Then the surface of the negatively biased electrode pushes away the particles. When another surface in the plasma region is biased less negatively or even positively with respect to the floating potential, the grains move toward these surfaces. In order to quantitatively investigate this effect, a floating glass substrate has been situated perpendicularly to an rf-driven electrode in a typical dusty plasma environment, see Figs. 1 and 2.

### 2.1.1. Experimental setup PULVA1

The experiments were performed in the PULVA1 reactor (Fig.1) [31]. It consists of a vacuum chamber with *40 cm* diameter which is pumped by a turbo-molecular pump (*260 l/s*) backed by a membrane pump. The residual gas pressure is *$10^{-4}$ Pa*. The working gas (Ar) is introduced into the vacuum chamber by a flow controller. The discharge is driven at *13.56 MHz* by a radio-frequency (rf) generator coupled to the bottom electrode by a matching network. The bottom electrode of the asymmetric rf discharge has a diameter of *13 cm* and is situated near the centre of the chamber; the chamber's wall serves as the other electrode. An adjustable butterfly valve is mounted between pump and vessel for controlling the gas pressure. Measurements were carried out at pressures of *1 ...10 Pa* and at rf powers of *3...50 W*. The microscopic test particles (SiO$_2$, *1μm*) which are levitated in the sheath in front of the driven electrode were illuminated by a laser fan (*532nm*) and observed by a CCD camera.

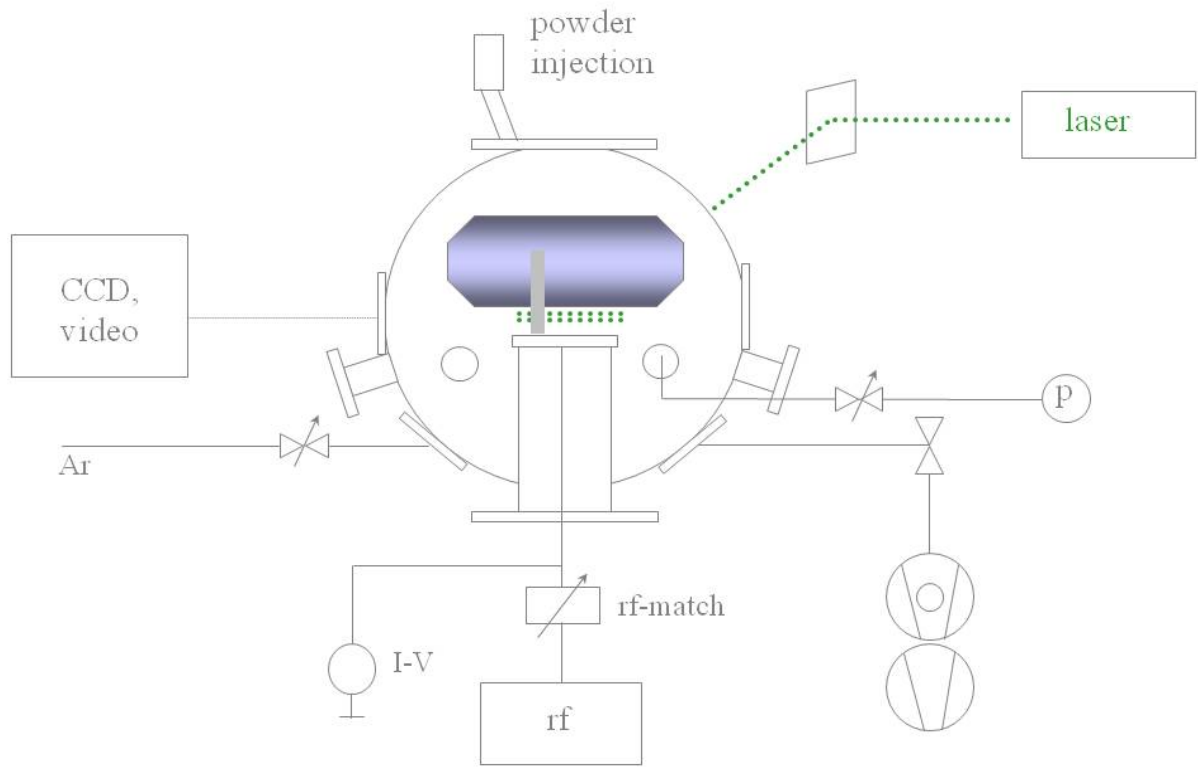

*Fig.1: Schematic view of the experimental set-up PULVA1.*

### 2.1.2. Results and discussion of the experiments in PULVA1

The superposition of different forces, caused by the self-biased electrode and the floating glass wall, influences the shape of the particle cloud. The glass wall, which extends into the plasma and which is on floating potential, causes an attractive force on the particles towards its surface. As a result, the surrounding sheath and the trapped particles show a kind of "wetting behaviour". This comparison with the capillarity of fluids springs into mind when looking at Fig.2 under this point of view [32].

At the photograph 1 in Fig.2 the common situation for low discharge power (*3W*) can be seen. The rf electrode is at the bottom and a lot of particles lie on it. At the left part of the photograph the glass box stays on the electrode. Above the electrode a nearly 2D particle cloud is levitated by the balance between gravitation and electrostatic force due to the electric field induced by the dc self-bias of the rf electrode. If the discharge power and, hence, the potential and field strength in front of the electrode changes with respect to the potential and field at the glass wall (photographs 2-6 in Fig.2), the balance between gravitation and the superposed electrostatic field forces shifts, too. Since the particles are only confined in regions where the force balance is fulfilled the changing in the shape of the particle cloud and its "wetting behaviour" can be explained qualitatively.

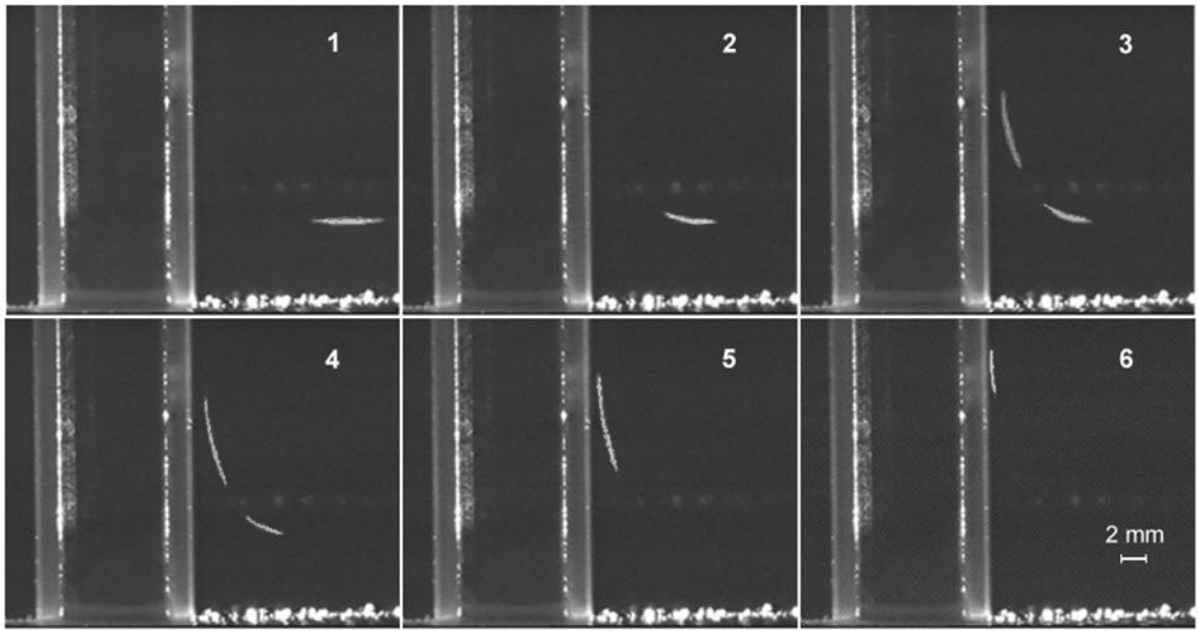

*Fig.2: Wetting of a glass wall by particles due to superposition of electrostatic fields (Ar plasma, p=6Pa, P=3, 6, 8, 10, 15, 44W from photograph 1 to 6).*

For a quantitative description of the experimental observation we balance the different forces acting on a particle, as depicted in Fig.3. Let us consider a simplified situation in a plane perpendicular to the rf electrode and glass box. In this cross section an element *dx* of the rf-electrode with the linear charge density $\sigma^1$ causes an electrostatic field force $dF^1$ in a distance $r_1$ on the particle with charge *Q*. Furthermore, an element *dy* of the floating glass wall with the linear charge density $\sigma^2$ causes an electrostatic field force $dF^2$ in a distance $r_2$ on the particle.

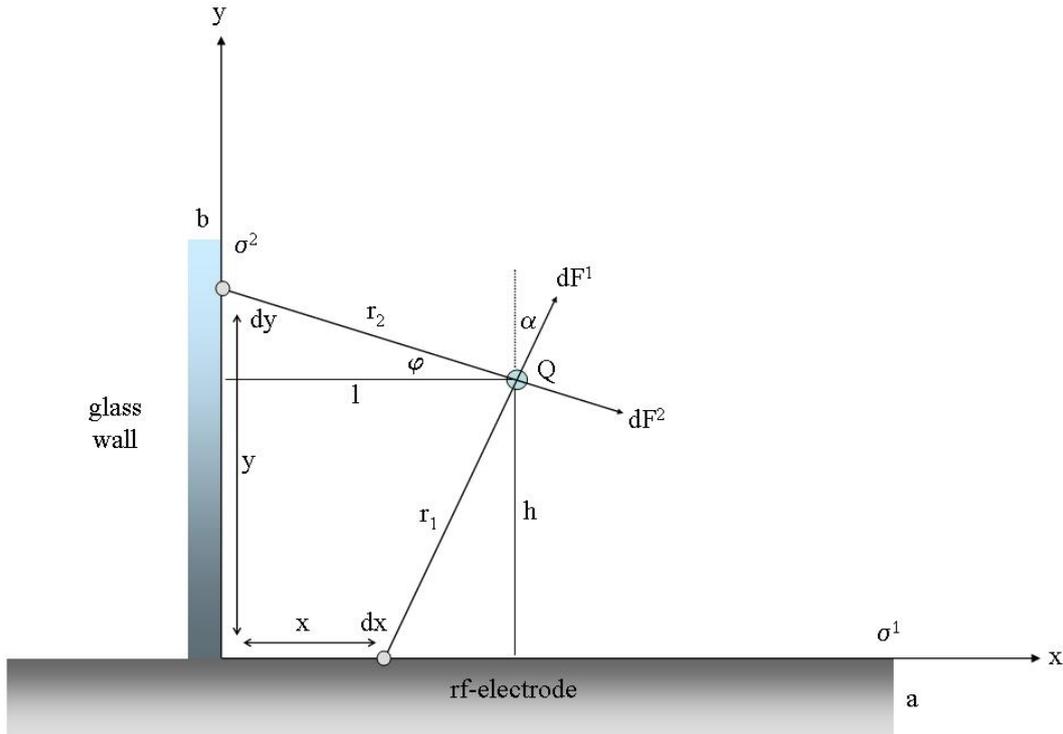

*Fig.3: Electrostatic forces acting on a charged particle (Q)*
*due to the charge density distributions ($\sigma$), e.g. potentials, at rf-electrode and glass wall.*

The components of the forces $dF^1$ into x- and y-direction are given by

$$dF_x^1 = \frac{Q\sigma^1 dx}{r_1^2} \cdot \sin\alpha \quad , \quad dF_y^1 = \frac{Q\sigma^1 dx}{r_1^2} \cdot \cos\alpha \quad , \tag{1}$$

and the components of $dF^2$ are given by

$$dF_x^2 = \frac{Q\sigma^2 dy}{r_2^2} \cdot \cos\varphi \quad , \quad dF_y^2 = \frac{Q\sigma^2 dy}{r_2^2} \cdot \sin\varphi \quad . \tag{2}$$

By taking into account the geometry of the experiment as shown in Fig.3 the electrostatic forces ($F_x^1$, $F_y^1$) between an integrated line of the rf-electrode and the particle in x- and y-direction as well as the electrostatic force ($F_x^2$, $F_y^2$) between an integrated line of the glass wall and the particle are obtained.

Finally, the components of the forces in x- and y-direction have to be summed for the equilibrium case

$$0 = \sum F,$$
$$0 = -F_x^1 + F_x^2,$$
$$0 = F_y^1 - F_y^2 - mg,$$
(3)

which results in the following equations:

$$\frac{Q\sigma^2}{l}\left[\frac{b-h}{\left[l^2+(b-h)^2\right]^{1/2}}+\frac{h}{\left[l^2+h^2\right]^{1/2}}\right]-Q\sigma^1\left[\frac{1}{\left[h^2+(l-a)^2\right]^{1/2}}-\frac{1}{\left[l^2+h^2\right]^{1/2}}\right]=0,\quad(4)$$

$$-\frac{Q\sigma^1}{h}\left[\frac{l-a}{\left[h^2+(l-a)^2\right]^{1/2}}-\frac{l}{\left[l^2+h^2\right]^{1/2}}\right]+Q\sigma^2\left[\frac{1}{\left[l^2+(b-h)^2\right]^{1/2}}-\frac{1}{\left[l^2+h^2\right]^{1/2}}\right]=mg.\quad(5)$$

Here $x, y$ are the coordinates of the surface elements; $h, l$ are the distances of the particle from the electrode and the glass wall; $a, b$ are the length of electrode and height of the wall, respectively.

Assuming typical values for the particle charge and the charge distribution along the rf-electrode and the glass wall, respectively, one obtains the equilibrium positions for the particles in the neighbourhood of the floating glass wall. In Fig.4 the results of the model are compared with the experimental observation for similar discharge conditions as shown in Fig.2. The motion of the particles to new equilibrium position and the formation of a new shape of the particle cloud can be correctly described by the model.

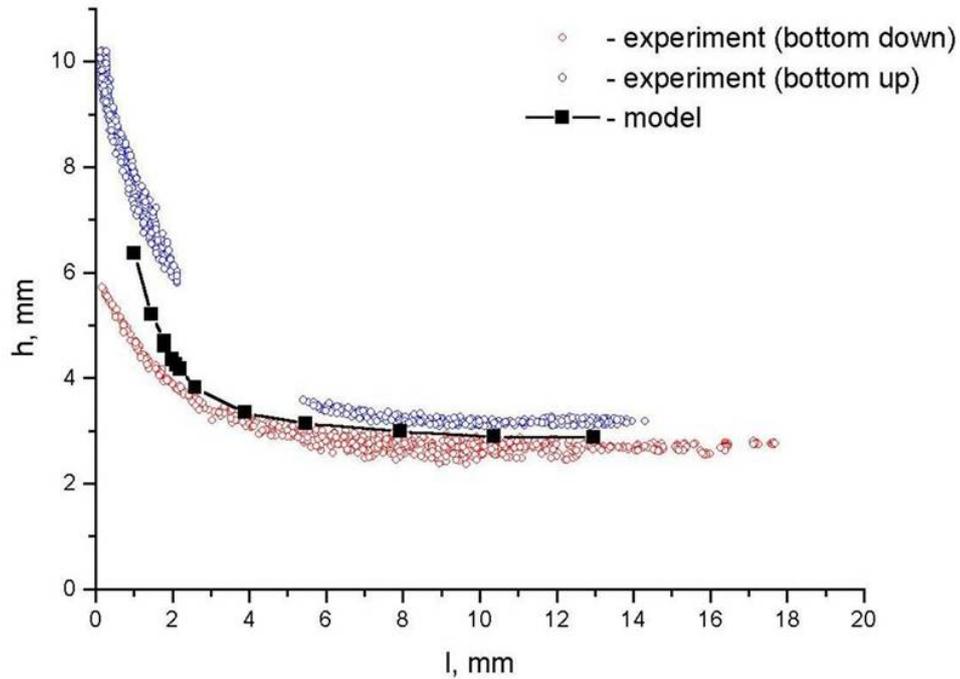

*Fig.4: Comparison of measured particle positions (SiO$_2$, 1µm) and model, "bottom down" means that the bottom of the glass box was at the electrode.*

Of course, for the calculation of the particle positions one has to make assumptions on the particle charge as well as on the linear charge densities of the surfaces. However, because these assumptions result in calculated values which are in good agreement with the experimental observations – this is a chance to get information on the charge distributions.

## 2.2. Particle behaviour in front of dc-biased surfaces

In order to influence the particles and to simulate different electrostatic surface conditions a segmented Adaptive Electrode (AE) [33,34] has been used as essential part of the experimental setup *PULVA-INP*, see Fig.5. This device is well suited for these investigations because the sheath structure can be manipulated locally. Furthermore, the influence of additional plasma sources (e.g. external ion beam source or sputter magnetron) on the behaviour of micro-particles in the rf-plasma can be investigated [30,35].

### 2.2.1. Experimental setup PULVA-INP

Again, a typical asymmetric, capacitively coupled rf-plasma in argon (*0.1…100 Pa*) is employed to charge the particles which are spherical melamine-formaldehyde (MF) particles of *0.5, 1, 5*, and *10 μm* in diameter. The reactor possesses an upper powered electrode (PE) and a lower adaptive electrode (AE) with 101 square pixels which can be biased independently by dc voltages as detailed below.

The rf-power (*5…100W*) is supplied by the upper electrode at a frequency of *13.56 MHz* and amplitude up to *2500 V*. In dependence on the discharge conditions we measured electron densities of $10^9...10^{11}$ *cm*$^{-3}$, electron temperatures of *0.8…2.8 eV*, and plasma potentials with respect to ground of *20...30 V* for the pristine plasma [37]. The plasma is bounded to the surrounding surfaces by the self-organizing structure of the sheath which has a characteristic potential slope and charge carrier profile. This region in front of the AE has been probed by the micro particles.

The particles are illuminated by a laser fan (*532 nm*), their positions and movements have been observed by a fast CCD camera and video recording which we employed to investigate the distribution of the powder particles in the plasma sheath, see Fig.5.

At low pressure (*p < 1 Pa*) the sheath is collision-less, whereas at higher pressure (*p > 10Pa*) it is dominated by collisions. Most experiments have been carried out just in the transition regime which is of interest for plasma processing. The transition behaviour can be clearly observed in the ion energy distribution (IEDF) at the grounded surface [13,37]. The maximum ion energy varies between nearly thermal energy due to thermalization and charge exchange by collisions at high pressure and nearly plasma potential (*~25eV*) due to collision-less transfer from plasma bulk to the surface.

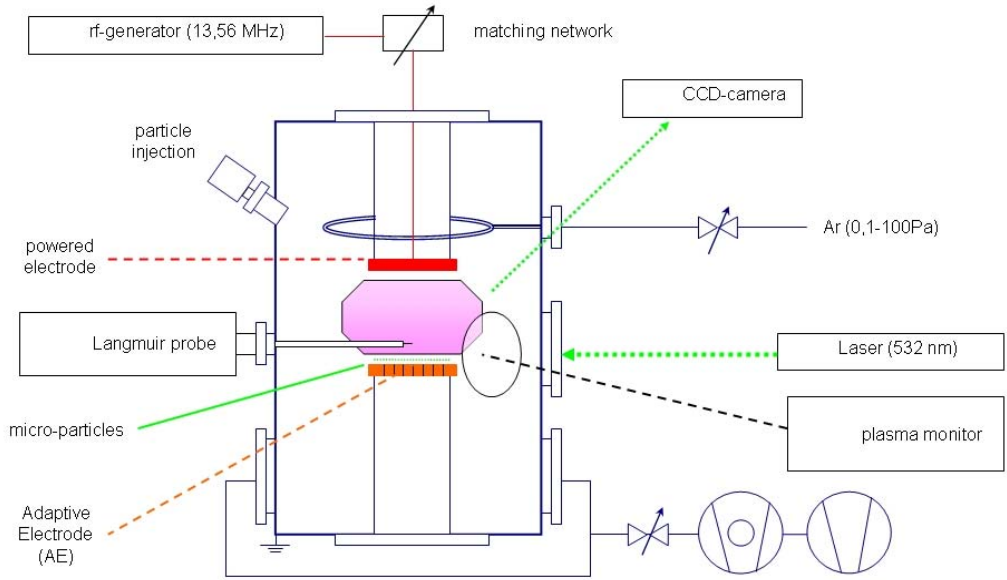

*Fig.5: Schematic of the set-up PULVA-INP. The central part is the AE.*

If the boundary wall (electrode, substrate) is without any external potential (e.g. floating), the electron and ion fluxes towards the surface are equal and the floating potential $V_{fl}$ reflects the internal plasma properties. An external biasing of selected surface elements (pixels) results in a local change of the sheath and the plasma. This idea is just the basis for the concept of the Adaptive Electrode (AE) [33,34] by which a spatial and temporal manipulation of the plasma sheath is possible. The 101 identical square electrode segments of the AE have a size of *7 x 7 mm²*. They are surrounded by 4 larger segments in order to fit the circular geometry of the planar electrode, see also left panel Fig.6. All 105 "electrode pixels" can be biased individually or in groups by an external *±100V* dc-voltage or ac-voltage (sinus, square, triangle shape) up to a frequency of *50 Hz* and any phase. In addition, for 3 segments of the AE also an rf-power supply (*13.56 MHz*) up to *4 W* is possible. The whole ensemble of electrode segments is surrounded by a ring electrode and a ground shield.

### 2.2.2. Results and discussion of the experiments for dc-biased pixels

Results in the literature on the electric field in the plasma sheath indicate a basically linear decrease with distance from the wall towards the bulk plasma [14]. As the plasma potential is positive with respect to grounded surfaces the electric field is directed towards the wall. Applying a local bias to some pixels of the AE influences the potential structure in the sheath and thus locally changes direction and magnitude of the electric field. In this way we can tailor confinement potentials for particles which are levitating above the AE. The particles will adjust their position such, that gravitation and electrostatic force in vertical direction balance each other, defining an equilibrium position. Around this equilibrium plane, the particles may oscillate (harmonically), but will be more or less confined to this region. It is clear that this equilibrium position will depend on the actual value of bias potential which is applied to the corresponding pixel of the AE. The potential differences at the surface induce additional forces in horizontal direction which are strongest at the interface between two pixels of different bias. Using a cloud of probe particles, which will arrange such that their potential energy is minimised, the spatial variation of the potential can be mapped. The shape of this equipotential surface is influenced by several factors. Despite step-like potential differences between adjacent pixels, the potential distribution above the AE will be smooth due to the mediating effect of the plasma. Depending on the plasma parameters, this smoothing effect will be more or less pronounced, alike the shielding of the particle charge.

Reproducing the experimental findings in a numerical simulation strongly depends on a successful inclusion of those aspects into the model. Treating the dust particles as probes which move in a fixed external potential is not fully correct, as the particles locally influence the surrounding plasma.

However, this influence is shielded on the scale of a few (one or two) Debye lengths and for larger distances from the particle this approximation is reasonable. For shorter distances corrections will be necessary. In our numerical simulation we statically account for the shielding of the particle charge by substituting the pure Coulomb interaction between different dust particles by a potential of Yukawa-type [36]. For the potential transition above different pixels, we assume a *tanh*-like shape, whose slope depends on the plasma parameters. In vertical direction, the particle motion is not restricted to the equilibrium plane, but confined by a harmonic potential around it.

Basing on these physical assumptions, the numerical simulation of the particle cloud is quite straight forward. Using molecular dynamics (MD) techniques, we solve the classical equations of motion for N particles,

$$m_k \ddot{\vec{r}}_k = \vec{F}_k \qquad , \qquad k = 1, \ldots, N \,. \tag{6}$$

Here $m_k$ denotes the mass of particle $k$, $\vec{r}_k$ its position and $\vec{F}_k$ is the total force acting on it, which consists of three contributions:

$$\vec{F}_k = m_k \vec{g} + q_k \vec{E}_k^{\mathrm{P}} + \sum_{j \neq k} \frac{q_k q_j}{|\vec{r}_{kj}|^3} e^{-\kappa |\vec{r}_{kj}|} \vec{r}_{kj} \,. \tag{7}$$

In the first term $\vec{g}$ denotes the acceleration of gravity. The second term is the electric force due to the electric field of the plasma $\vec{E}^p$ and $q_k$ is the charge of the particle. The Yukawa interaction between the particles in the third term is characterised by the shielding constant κ and the relative position of the particles, $\vec{r}_{kj} = \vec{r}_k - \vec{r}_j$. For κ = 0, this term is just the (unscreened) Coulomb force.

To lowest order approximation we assume the monodisperse dust particles to be equally charged. The presence of other particles in the vicinity of a dust grain, however, influences the local plasma properties, e.g., ion and electron densities. As the particle charge reflects the local charge equilibrium between electron and ion currents towards the particle, the altered environment will also lead to a different dust charge. For the moment, we neglect this dependency to get a first crude estimate of the potential structure above the AE.

Starting from a randomly chosen set of initial conditions, we calculate the trajectory of each particle. If we include an additional friction term and, thus, dissipate kinetic energy, the particles will relax into a configuration that minimises the total potential energy. Clearly, the energy obtained in this way is not a true global minimum as for such a minimisation process a more sophisticated energy reduction is required (e.g. the simulated annealing process in Monte-Carlo simulations [39]). But here the global minimum is of minor importance, as it is improbable that our classical system at finite temperatures will reach this state. All we need to have is a set of probable configurations with low energy (local minima) into which the system relaxes for different sets of initial conditions. This gives us the necessary information about the equipotential surfaces. The occurrence of a friction term can also be motivated from a physical point of view. Moving through the background plasma, the particles collide stochastically with ions and neutral Ar atoms, loosing constantly momentum and energy. Thus such a term accounts for ion and neutral drag on the dust grains. In the right panel of Fig. 6 we show the result of a typical MD run simulating the experimental setup in the left panel. Despite the simplicity of the assumptions on shielding and smoothing of the potential differences due to the plasma, the accordance between simulation and experiment is remarkably good. Critically contrasting the two results, we remark differences in the corners of the structure. Apparently, the used parameters are not capable to reproduce the sharp edges in the bars of the 'Φ' seen in the experiment (left panel of Fig. 6). Here the assumption of a mere superposition of horizontal *tanh*-profiles for the potential is too simplistic. We will later present a full three-dimensional PIC simulation which enables us to study the true potential structure above the pixels.

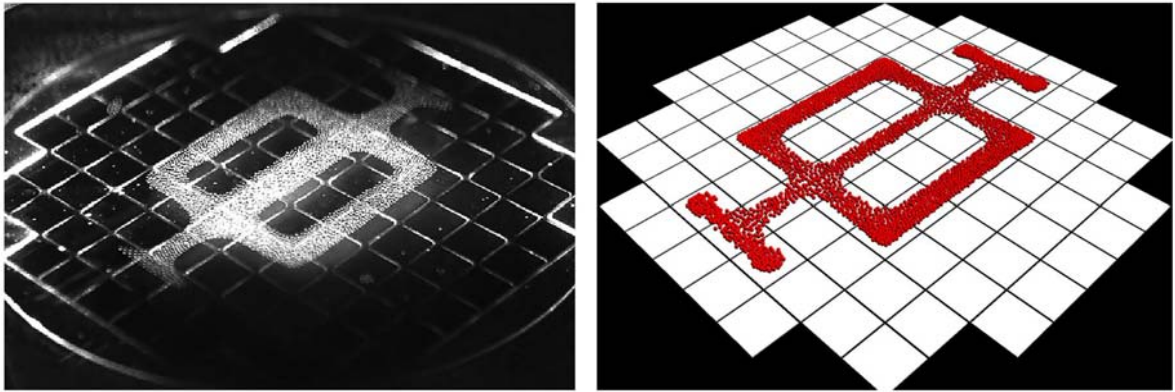

*Fig.6: Left: Pattern of particles (MF, 9,6μm) which are charged in an the rf-plasma and levitated in front of the AE due to different pixel-biasing. Right: Simulation results for the same system using 3000 particles.*

For a more detailed examination of the electric field structure in front of a wall, here the AE, we need to know the charge of the probe particles. In the literature, several methods for determining the particle charge can be found; for a summary see e.g. [40-42]. One of the most common methods is the excitation of trapped probe particles to oscillate around their equilibrium positions by applying an external low-frequency voltage [7,15,40].

In the experimental setup, a single MF-particle is confined above the centre pixel of the AE and its equilibrium position $z_0$ is measured. An additional sinusoidal voltage applied to the centre pixel, causes the particle to oscillate around $z_0$. Recording the oscillation amplitudes for different driving frequencies allows for a determination of its resonance frequency $\omega_0$.

In Fig. 7 we present results on equilibrium position and resonance frequency in dependence on the discharge conditions, e.g. neutral gas pressures. Using particles of different size, the corresponding equilibrium positions cover a wide range of the sheath and allow for a thorough characterisation of the electric field in the sheath and the particle charge.

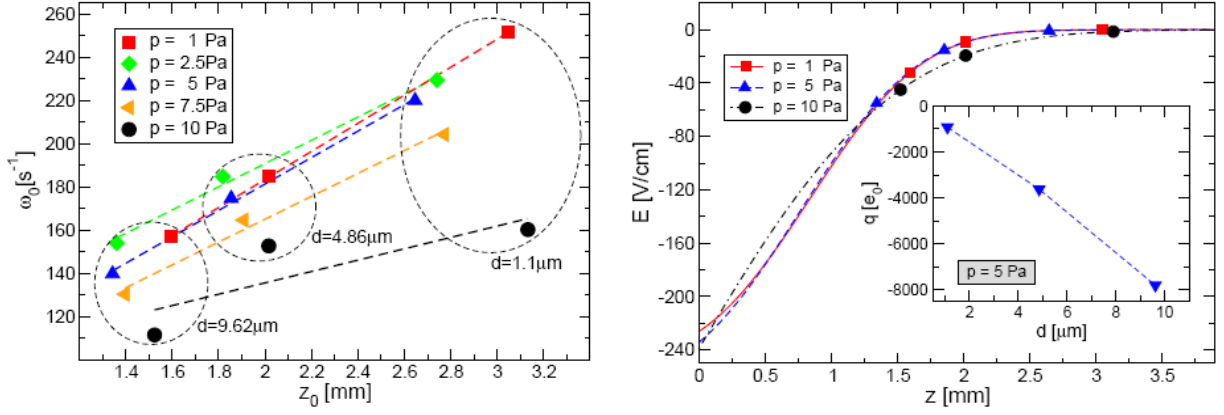

*Fig.7: Left: Experimental results for probe particles in the sheath in front of the AE: Relation between resonance frequency ω₀ and equilibrium position z₀ for different particle diameters d and neutral gas pressures p. Right: Calculated electric field E as a function of distance z from the AE for different neutral gas pressures p. Inset: Particle charge q as a function of particle diameter d for p = 5 Pa. These results are deduced from the data in the left panel.*

We may exploit the obtained relation between $\omega_0$ and $z_0$ to determine the particle charge and electric field structure in the sheath. To this end, we first neglect both drag forces and phoresis effects and describe the system as a driven harmonic oscillator. The equilibrium position $z_0$ of the (negatively) charged dust particle (charge $q(z) = -Z(z)e_0$) is determined by $-q(z_0)\vec{E}(z_0) = m\vec{g}$, where the number $Z(z_0)$ of elementary charges $e_0$ on the particle will depend on its vertical position $z$, $m$ is the mass of the particle. With $\vec{E}(z) = E(z)\vec{e}_z$ and $\vec{g}(z) = g(z)\vec{e}_z$ we get $-e_0 Z(z_0)E(z_0) = mg$. Upon applying a low-frequency voltage, the oscillations of the particle around $z_0$ are harmonic [25] provided the amplitudes are not too large. Within the harmonic oscillator model the dust charge can thus be considered constant in the vicinity of the equilibrium position.

The resonance frequency of the particle at position $z_0$ is given by

$$\omega_0^2(z_0) = Z(z_0)\frac{e_0}{m}\frac{dE(z)}{dz}\bigg|_{z_0} = -\frac{g}{E(z_0)}\frac{dE(z)}{dz}\bigg|_{z_0}, \qquad (8)$$

where we used the equilibrium condition, to eliminate *m* in the second expression. This differential equation for the electric field can be solved by separation. Formal integration yields

$$E(z) = E(0) \exp\left(-\frac{1}{g}\int_0^z \omega_0^2(\zeta)\, d\zeta\right).$$
(9)

Equating the (negative) integral over the electric field across the sheath with the sheath voltage fixes the value of *E(0)* at the surface of the AE. For a further evaluation we need to know the relation between resonance frequency and equilibrium position $\omega_0(z)$ throughout the whole sheath. Experimental data (Fig. 7, left) suggests a linear behaviour over a wide range of the sheath for low and moderate pressures (*p < 7.5Pa*). For *p = 10Pa*, the available data and the assumption of a linear relation between $\omega_0$ and *z* agree only poorly. Unfortunately, we are not able to obtain experimental data in the very close vicinity of the AE due to surface attachment of the particles. Using a linear ansatz, $\omega_0 = a_0 + a_1 z$, in eq.(9) we obtain the electric field,

$$E(z) = E(0) \exp\left(-\frac{a_1^2}{3g}z^3 - \frac{a_0 a_1}{g}z^2 - \frac{a_0^2}{g}z\right),$$
(10)

in the sheath in front of the AE, which we show in the right panel of Fig. 7 for different pressures. Utilising the knowledge about the electric field, we can directly obtain the particle charge at its equilibrium position (inset in right panel of Fig. 7).

To check the reliability of the above results and the validity of the underlying assumptions, we compare the obtained electric field in the sheath (eq.10) with results from independent PIC simulations in the left panel of Fig. 8. Details on the PIC code are given in section 2.3. Taking these fields as fixed input, we are able to determine $z_0$ and $\omega_0$ of the particles as a function of the particle charge by the MD method described above (right panel of Fig. 8).

A possible explanation for the discrepancy between the results for the electric fields in Figs. 7 and 8 is the following: Experimentally determining the resonance frequency requires finite amplitudes. During the oscillations the particles exceed the range of validity of the local harmonic approximation. For such amplitudes also the particle charge will not be constant anymore which is presumably the main effect. The particles adapt to their local environment much faster ($\mu s$) than they move through the sheath [38]. This adiabatic adjustment of the particle charge is not contained in the simple harmonic oscillator model.

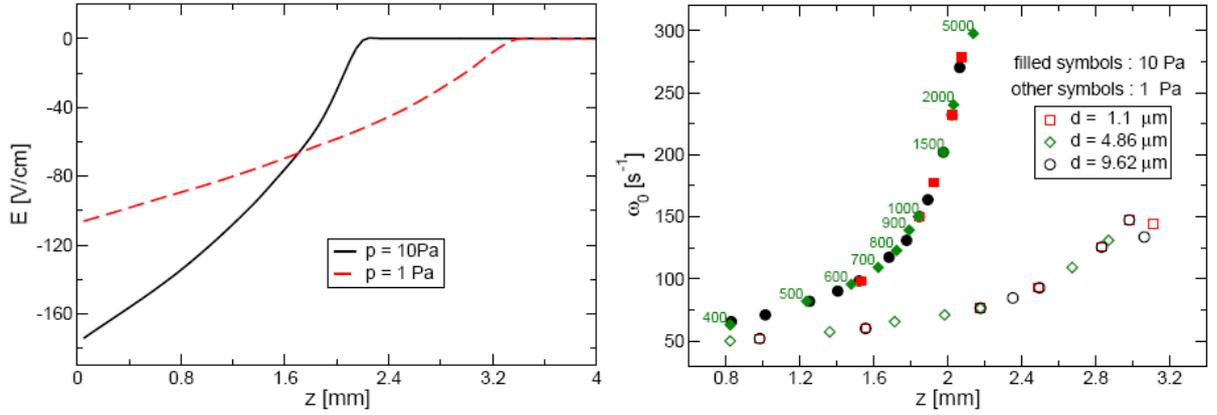

*Fig.8: Left: Results from PIC simulations for the electric field E in front of the AE for different neutral gas pressures p. Right: Numerically (MD) obtained resonance frequencies and equilibrium positions for different particle charges q and the electric field from the left panel. For p = 10 Pa the charge on the particles with diameter d = 4.86μm is given for each data point in units of $-e_0$.*

Overall, i.e. also in front of grounded or additionally biased surfaces, we may experimentally determine the electric field structure of the sheath by means of charged micro-particle probes. The construction of the AE also allows for experiments on the sinking and oscillatory behaviour of individual particles caused by a change of the local bias potential as already proposed by other authors [15,43-45]. To this end, a single particle (MF, *9,6μm*) is trapped above the centre pixel (E5) of the AE by the confining potential shown in Fig.9.

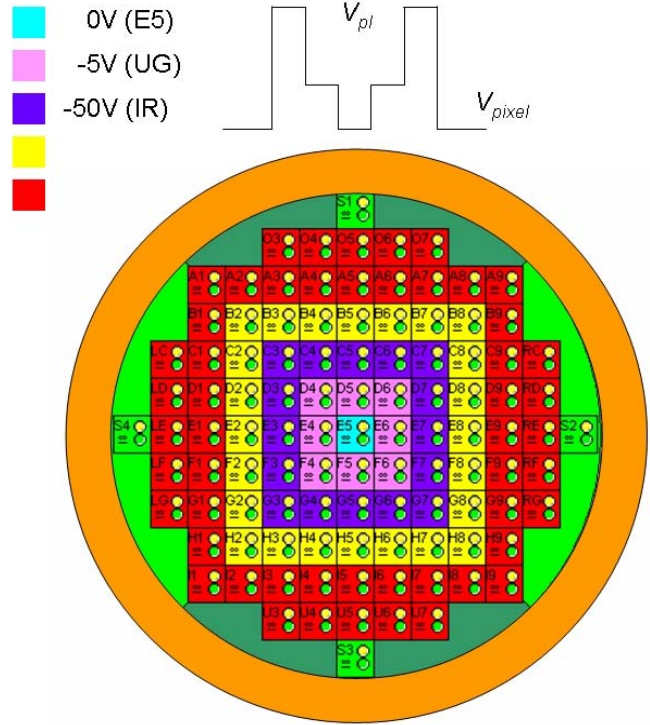

*Fig.9: Structure of the AE. The centre pixel (E5) can be biased in order to influence the sheath for particle trapping, here it is on ground. The surrounding pixels (UG, IR) are biased as indicated for the particle confinement in the potential trap. The surrounding segments (yellow, red, green) are at ground potential.*

The net force on the particle in vertical direction is given by

$$F(z) = F_{el}(z) + F_n(z) - F_g - F_{ion}(z) \qquad (12)$$

with the electrostatic force $F_{el}$, the neutral drag force $F_n$, the gravitational force $F_g$ and the ion drag force $F_{ion}$. The net force vanishes at the equilibrium position $z_0$. If the confining plasma is switched off, and the pixel E5 is on ground potential ($V_{bias}=0$), the forces $F_{el}$ and $F_{ion}$ become zero. In the vacuum case the simple free-fall condition for the particle is valid. However, taking into account the neutral drag force $F_n = -\beta \dot{z}$ the force balance is given by

$$F(z) = m\ddot{z} = -mg - \beta\dot{z} \quad . \tag{13}$$

Solution of this differential equation results in

$$z(t) = z_0 - \frac{mg}{\beta}\left(t - \frac{m}{\beta}\left(1 - e^{-\frac{\beta}{m}t}\right)\right) \quad . \tag{14}$$

The diagrams for the free fall and the real fall of the particles after switching-off the plasma are plotted in Fig.10. The motion begins at the equilibrium position $z_0$ where the particle is trapped. The sheath width in front of the AE (pixel E5) is about *3mm* and the equilibrium position at about *1,8mm*. From the experimental data shown in Fig.10 (circles) the damping constant $\beta$ by the gas friction can be estimated to about *2 $10^{-11}$kg/s* (solid line). For comparison, the dashed line in Fig.10 shows a particle trajectory in absence of any friction force (free fall).

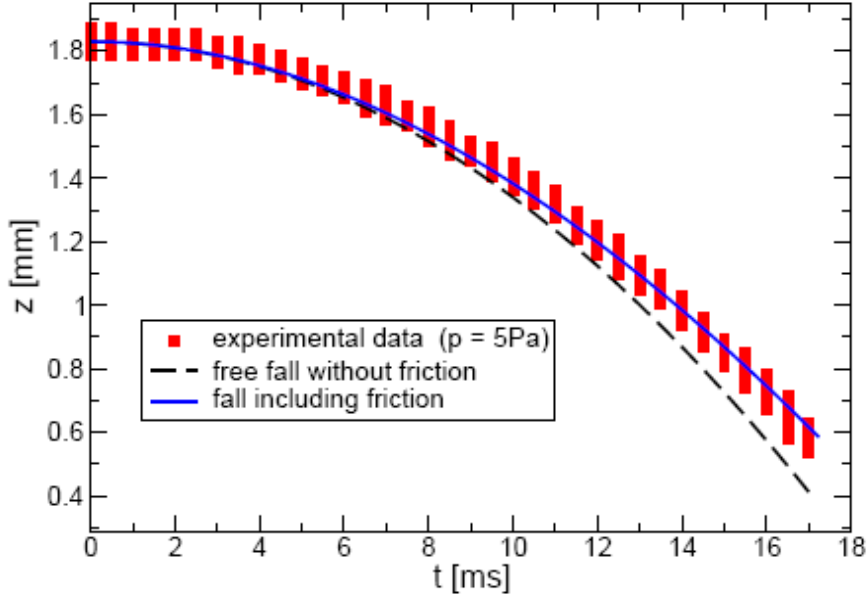

*Fig.10: Particle behaviour after switching-off the plasma.*

Keeping the plasma operational, we may also investigate the particle behavior caused by sudden changes in the sheath due to additional bias voltages on some pixels. Compared to the previous experiment, the description here is more involved, but it is also promising in view of gaining information on particle charge and field distribution. We start again from the experimental configuration shown in Fig. 9 and confine a single particle above pixel E5 with no additional bias on this pixel. Then applying a negative bias voltage $V_{bias}$ causes the particle to leave its initial equilibrium position. Depending on $V_{bias}$ we may distinguish two cases: Up to some critical value of the bias potential, the particle is pushed away from the electrode, and reaches another equilibrium position to which it relaxes after some oscillations around it. Above a critical $V_{bias}$ (about *-30V*), the behaviour changes fundamentally. In this case, the particle falls downwards and finally hits the AE. To explain these different behaviours consider the following: The equilibrium position of a particle is determined by the balance of gravitation and electrical force, which depends on the local electrical field as well as on the actual particle charge. A more negative bias voltage broadens the plasma sheath within a few rf cycles, reducing both the electron and ion density. This effect is drastically more pronounced for the electrons than for the ions. By construction, the biasing of the AE should only have a negligible effect on the bulk plasma, leaving the plasma potential unaltered. In the sheath, the electric field is enhanced due to the increased potential difference which surpasses the effect of the sheath broadening. An increase of the electric field strength implies, that a particle rests at its initial position, if its previous charge is decreased. Highly charged, the particle will move towards the bulk plasma as the electric force dominates the gravitation (in the opposite case it will move towards the AE). Depending on the electron and ion densities in the vicinity of the particle, the particle charge will adapt according to its local surrounding.

As compared to the particle dynamics, those recharging processes take place on a much faster time scale. This drastic change leads to such a fast loss of particle charge that the electric force is never able to compensate gravitation and the particle drops onto the AE. The experimental data indicates even an additional acceleration towards the AE, as compared to the free fall of an uncharged particle. For an explanation of this additional downward acceleration, the most obvious candidate is the electric force, provided we allow the particle to acquire a positive charge.

It is clear, that the particle charge reflects the local balance of electron and ion density. Applying a negative bias, only electrons in the high energy tail of the electron energy distribution function may penetrate into the enlarged sheath in front of the biased pixel. This leads to a further reduction of the local electron density while the global plasma parameters of the discharge and the global charge balance are unaffected. Locally, the sheath in the close vicinity of the biased pixel is completely deprived of electrons. In such a region, a dust particle is exclusively surrounded by ions. Ion-dust collisions will reduce the negative dust charge and eventually also a positive charging of the dust is possible. Positive particle charges in the afterglow of a plasma discharge have been reported in the literature [46-48]. However, to explain the observed behaviour exclusively by electric forces, unrealistic high charges would be necessary ($10^4 e_0$ for $V_{bias} = -30V$ and $1.7 \times 10^4 e_0$ for $V_{bias} = -50V$) from the beginning on. Despite a strongly reduced averaged electron density at the starting position of the particle, during some fraction of the rf period electrons still reach this region. Due to their higher mobility, this is sufficient to keep the particle charge negative, though reduced as compared to the initial value. This means, that a positive particle charging is only possible closer to the AE, in regions where the electrons are completely absent. Whether charges as high as mentioned above are possible, is however questionable. For the early stages of the fall, surely another explanation is necessary. Here the (up to now neglected) ion-drag force presents itself. The large negative bias enhances the ion current toward the AE-pixel and also increases the ion energies. By collisions, the ions transfer part of their momentum to the dust particle, explaining the additional acceleration. Closer to the electrode, the (higher energetic) ions may transfer more momentum, but their density is reduced. So there the charging effect will be dominant.

In total, the combination of the two forces allows for an explanation of the observed falling curves without the necessity of unrealistic large permanent positive dust charges. For lower bias, the effect of the ion-drag is also present, but as long as the ion and electron densities are still alike as before, it is dominated by the upward electric force.

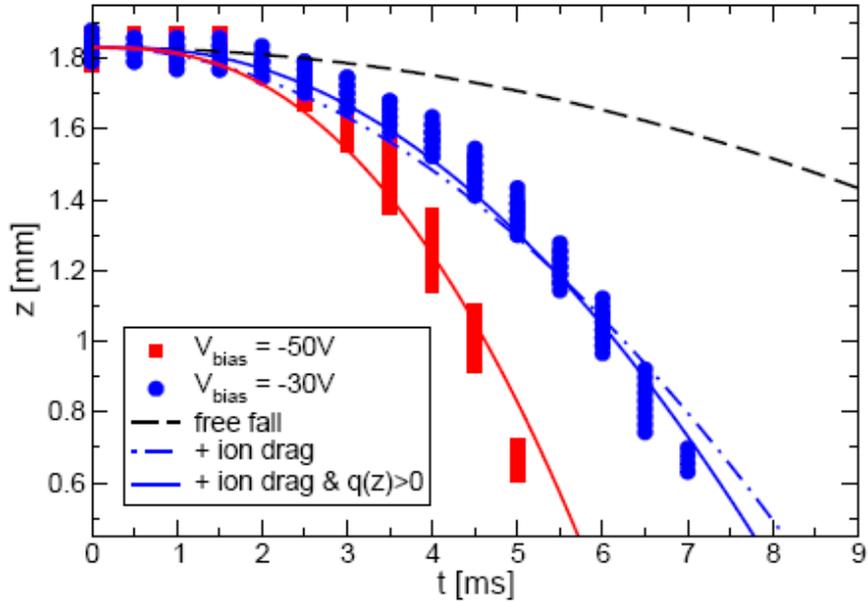

*Fig.11: Accelerated fall of a test particle after sudden application of a bias voltage of −30 and −50V. For comparison, the dashed line indicates the trajectory for a free fall. For the dashed-dotted and solid lines the ion drag force is also included. While for the dashed-dotted line the negative particle charge may only be reduced, for the solid line also positive charging is possible.*

**2.3. Particle behaviour in front of rf-biased surfaces**

In addition to the main power supply by the powered electrode, some pixels of the AE (here the central pixel (E5) again, see Fig.9) can be driven by an additional rf voltage with the same frequency (*13.56 MHz*) and phase to reach a local enhancement of the plasma. An example of such a plasma bubble or dome obtained in an argon plasma at a pressure of *7.5 Pa* is shown in Fig.12. The figure illustrates a pronounced local enhancement of the light emission on top of pixel E5. The lower part of the bright light represents a reflection of the light on the electrode surface. The margins of the square segments can be slightly recognized in Fig.12 again. The size of the light bubble grows with increasing power which is supplied through the pixel E5 up to *4 W*.

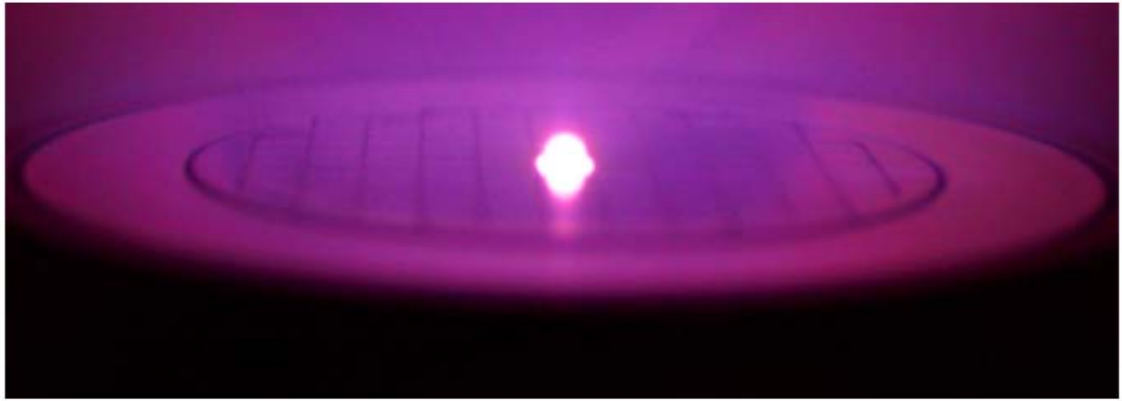

*Fig.12: Enhanced light emission (plasma bubble) above the pixel E5.*

*(E5: rf-power = 2.5W, UG: $V_{DC}$ = -15V, all other pixels: $V_{DC}$ = -50V )*

To describe the observed plasma enhancement a fluid model has been applied. In order to limit the numerical expenditure, the reactor was approximated by an axisymmetric geometry using concentric rings instead of the square pixel structures of the adaptive electrode. A similar model [49] has been used to describe the plasma in the same reactor, but operated using only with a dc voltage structure at the adaptive electrode.

The model consists of particle balance equations for the charge carriers, the electron energy balance equation and Poisson's equation. The transport of the charged particles is described by their particle balance equations

$$\frac{\partial n_\alpha}{\partial t} + \nabla \cdot \vec{\Gamma}_\alpha = S, \ \alpha = e,i \quad , \tag{15}$$

where $n_e$ and $n_i$ denote the particle densities of electrons and positive argon ions, respectively. The source term *S* describes the electron impact ionization of the argon atoms in ground state.

The time- and space-dependent particle fluxes $\vec{\Gamma}_\alpha$ of the charge carriers are determined in drift-diffusion approximation $\vec{\Gamma}_\alpha = -\mu_\alpha n_e \vec{E} - D_e \nabla n_e$ using their mobilities $\mu_\alpha$ and diffusion coefficients $D_\alpha$. This description is chosen as a firts approach because it neglects the inertia of ions which may come into play at low pressure. The source term $S$ and the transport coefficients $\mu_e$ and $D_e$ depend on the velocity distribution function of the electrons and on cross sections of electron collisions with the argon atoms. In the frame of the current approach a Maxwell distribution is assumed for the electrons and their temperature $T_e$ is determined by solving the electron power balance

$$\frac{3}{2}\frac{\partial}{\partial t}(n_e k_b T_e) + \nabla \cdot \vec{\Gamma}_e^\varepsilon = -e_0 \vec{\Gamma}_e \cdot \vec{E} - n_e \sum_r N k_r \varepsilon_r \qquad (16)$$

which includes the elementary charge $e_0$, the energy flux

$$\vec{\Gamma}_e^\varepsilon = -\frac{5}{2}\mu_e(n_e k_b T_e)\vec{E} - \frac{5}{2}D_e \nabla(n_e k_b T_e) \qquad (16a)$$

and the energy loss $n_e \sum_r N k_r \varepsilon_r$ due to elastic and inelastic collisions of the electrons with the gas atoms. Here, $\varepsilon_r$ and $k_r$ denote the energy loss of the r-th collision type and the corresponding collision rate coefficient, respectively and $N$ is the density of the gas atoms. Furthermore, the Poisson equation

$$\nabla \cdot \vec{E} = \frac{e_0}{\varepsilon_0}(n_i - n_e) \qquad (17)$$

is solved to determine the electric potential $\phi$ and field $\vec{E} = -\nabla \phi$ in the discharge. This set of equations is completed by an equation which accounts for the outer circuit and yields the potential at the powered electrode as a function of time. A simple circuit with a capacitor $C_B$ in series connection with the powered electrode has been taken into account. Consequently, the relation between the potential at the rf generator $\phi_G$ and that at the powered electrode $\phi_P$ is described by the equation

$$C_B \frac{d(\phi_P - \phi_G)}{dt} = e_0 \int_A (\frac{\partial \vec{E}}{\partial t} + \vec{\Gamma}_i - \vec{\Gamma}_e) \cdot d\vec{A}$$ , (18)

where the integral at the right-hand side is to be calculated across the whole surface of the reactor. This equation allows to determine the self-bias voltage $\phi_{dc}$ which arises in an asymmetric rf discharge.

Fig.13 shows results of the model calculations. The electric potential in the active volume between the electrodes and in the adjacent vessel volume is represented in Fig.13a. At the adaptive electrode, three sections with different boundary conditions can be distinguished. The pixel E5 with a radius of *4mm* is followed by two rings ($R_1$ and $R_2$) which are *7.6* and *54.6mm* wide, respectively. The rings were supplied with dc voltages of *-15V* ($R_1$) and *-50V* ($R_2$), respectively. The pixel was driven with a dc bias of *+17V*. The amplitude of the rf voltage $U_{E5}$ was chosen equal to *100V* to obtain in the volume $V_{E5}$ surrounding E5 nearly the same rf-power consumption $P_{E5}$ of *2.5W* as in the experiment. This was checked by integrating according to

$$P = \frac{e_0}{\tau} \int_0^\tau \int (\vec{\Gamma}_i - \vec{\Gamma}_e) \vec{E} dV dt$$ (19)

over the volume $V_{E5}$. The main electrode at *z=10cm* is powered with an rf voltage of *332V* and a bias of *-283V* which corresponds to a total rf-power of *12.5W* when calculating the integral (eq.19) over the total volume. Figure 13a presents the potential distribution at the phase instance $\pi$ by a contour plot and additionally the potential along the centre of the discharge at the instances $\pi/2$ and $3\pi/2$ by line plots. A strong modulation of the potential near the pixel E5 and the powered electrode can be observed. In the quasi-neutral bulk plasma the potential is nearly space-independent but oscillates between *31V* and *39V* in the centre.

Fig. 13b gives an enlarged representation of the averaged electric potential in the vicinity of pixel E5. The result for $U_{E5}$ *=100V* is compared with that obtained for $U_{E5}$ *=50V*. The increase of the potential in the bulk plasma qualitatively agrees with the experimental observations. A local maximum of the potential is formed in front of the electrode which moves towards the bulk plasma with increasing $U_{E5}$. At the bottom, the magnitude of the electric field for $U_{E5}=100V$ and its direction are shown as contour and arrow plot. The latter makes a field reversal obvious. The contour lines illustrate the constriction of the sheath around the pixel E5.

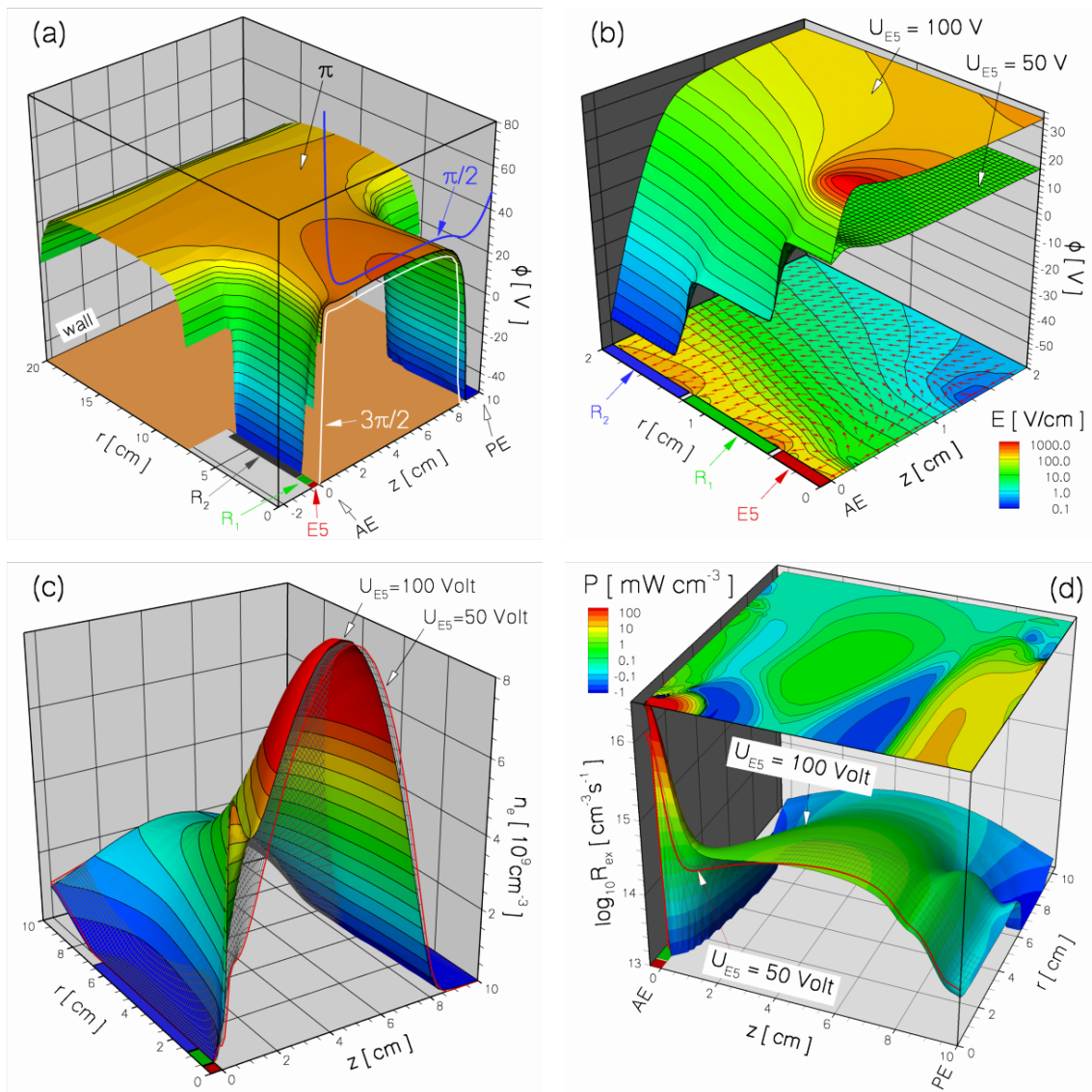

Fig.13: Plasma characteristics calculated for an argon plasma additionally excited by an rf voltage at the pixel E5: a) electric potential at phase instances $\pi/2$, $\pi$ and $3\pi/2$, b) period

*averaged electric potential (top) and field (bottom) in the vicinity of E5,  c)  averaged electron density, and  d) averaged excitation rate (bottom) and power density (top).*

Besides the broad maximum in the centre of the discharge a pronounced second maximum at a distance of about *5mm* from the pixel E5 is found for $U_{E5}$ *=100V*. This maximum is caused by the increased electric field and the resulting increased ionization and also secondary electron emission from the pixel in this region. Fig.13d displays at its bottom the rate of exciting collisions of electrons with argon atoms. Its pronounced maximum in front of E5 corresponds to the shape of the enhanced light emission of the dome shown in Fig.12. With increasing voltage the magnitude and spatial extension of the enhanced plasma region increases as observed in the experiment. At the top of Fig.13d the averaged power gain density of the electrons is presented for $U_{E5}$ *=100V* as a contour plot. The power gain density in front of the pixel E5 is even higher than near the main powered electrode (PE). This pronounced maximum is caused by the large electric field, electron density and correspondingly also large electron flux near the pixel E5.

As a complementary approach to the fluid description, we investigate this system also by a particle-in-cell (PIC) method. Instead of trying to describe the global plasma parameters in the whole reactor vessel, here we focus on local modifications of the plasma sheath caused by the AE. In contrast to the fluid approach within the PIC scheme we fully account for the (non-equilibrium) distribution functions for electrons and ions. This is of particular importance in the spatially inhomogeneous regions of the plasma sheath and for low pressures, where low collision rates inhibit the formation of a thermalized equilibrium distribution. For a general introduction of fundamentals of PIC see, for example, [50] and references therein. There is a variety of open source codes available on the Internet, e.g. the code suites from the PTSG [51]. These codes are on the one hand designed to be very flexible, which on the other hand limits their numerical performance. Therefore, we started from the pdp2-code [52,53] as a basis and developed our own code, matching exactly our requirements without additional overhead. Without going too much into numerical details, we only want to stress the main physical aspects of this code. Due to the low temperature and moderate driving voltage of the discharge only weak currents arise in the plasma. Thus the generated magnetic fields are negligible and as no external magnetic field is present, we may content ourselves with an electrostatic code. In a first step, we consider a vertical two dimensional cut through the reactor vessel, restricting the simulation volume to a rectangular area between the two electrodes. For the solution of the Poisson equation on this rectangular grid we use cyclic reduction [54] as it is implemented in FISHPACK [55]. We choose reflecting boundary conditions at the boundary in the centre of the discharge to reduce the number of grid points by a factor of two due to symmetry. If the opposite boundary is far enough away from the region of interest, e.g. the sheath in front of the biased pixel of the AE, its influence on the local characteristics is negligible. Hence, we are free in our choice and take Dirichlet boundary conditions. The influence of the AE onto the plasma is taken into account by self-consistently balancing the charge on the individual pixel, which we connect each to an external circuit with voltage source and capacitor. By these means, we are not only in the position to investigate static dc-bias voltages on the AE pixels, but also study the effect of ac-bias or rf voltages. The collisions of the charged species with the neutral background gas are treated on the basis of the null collision method [56,57].

Although the PIC method allows us study the characteristics of the discharge and the sheath with high time resolution, in view of micro-particles as probes only average quantities are of interest due to their slower dynamics. Fig.14 compares the influence of biased pixels on the averaged potential, as well as the electron and ion density in the sheath. It is obvious, that a negative bias voltage distorts the equipotential surfaces and can thus be used as a confining barrier potential for particles on an adjacent pixel. In order to resolve the complete spatial structure of the sheath, we extend our PIC code to three dimensions, resulting in a full 3d3v-code with the above specifications.

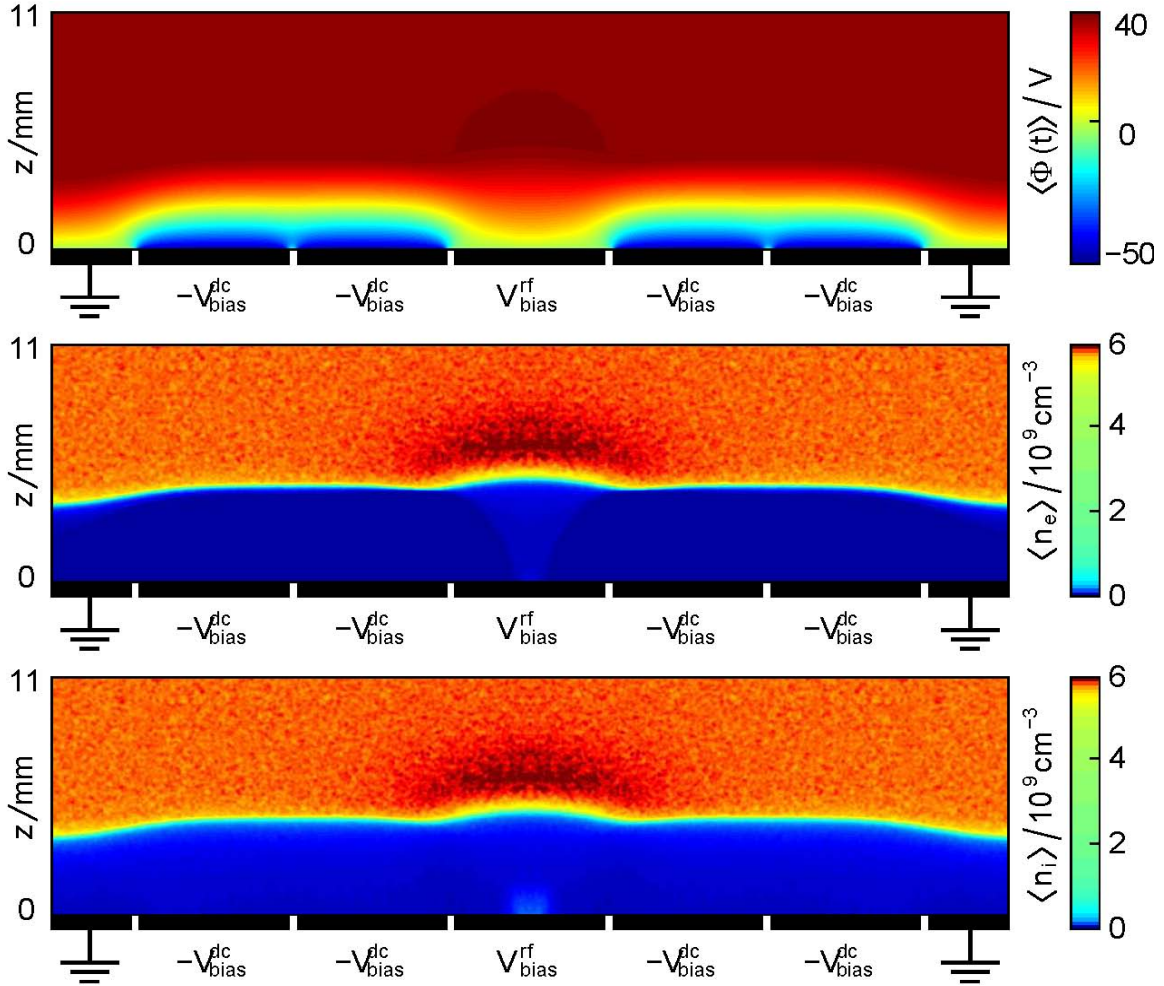

*Fig.14: Perturbation of the plasma sheath by an additional rf-biased pixel (centre) which is on $0 \pm 50V$. We show the influence on the time averaged potential (top) as well as on the time averaged electron (middle) and ion density (bottom). Results are based on a two dimensional PIC simulation for p = 10 Pa. The other pixels are biased by -50V.*

The solution of the Poisson equation is still performed using FISHPACK, and the used boundary conditions include now two perpendicular mirroring planes in the centre of the discharge. The desired benefit of these results is to get information about the potential structure above the corners of biased pixels, which can be compared against results as described above. In Fig.15 we show horizontal cuts through the sheath above a biased pixel.

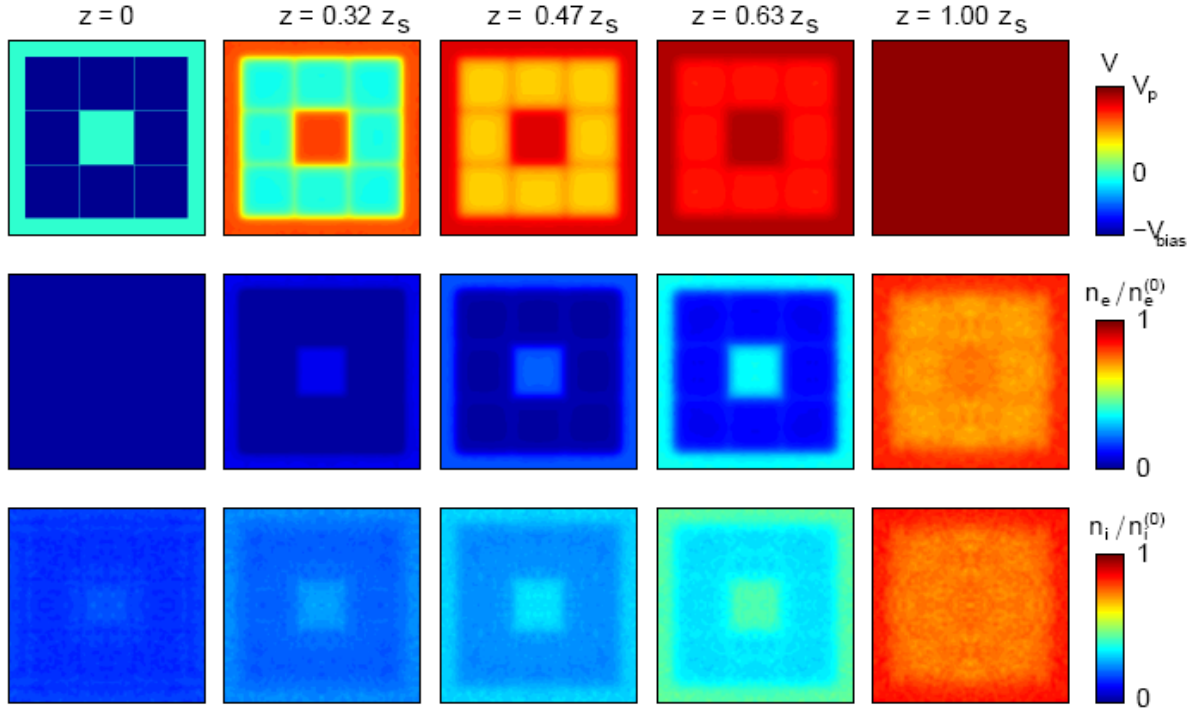

*Fig.15: Three-dimensional structure of the sheath in front of the AE. We show horizontal cuts through the sheath (thickness $z_S$) for different distances z above the AE for the time averaged potential as well as for electron and ion density, which are normalized to their bulk values $n_{e,i}^{(0)}$. The biasing of the AE pixels can be seen from the potential structure at z = 0, $V_{bias}$ = −20V, p = 10Pa.*

A very interesting phenomenon was observed by injecting test particles into the plasma which interact with the dome-like plasma of the rf biased centre pixel E5. The particle behaviour can be seen in Fig.16 and in the attached movie.

Some MF particles were confined in front of the pixel E5 and then the plasma rf power at this pixel was increased from *0* to *4W*. The potential conditions of the surrounding pixels are not changed. At the beginning, with increasing power the particle cloud moves slightly into the direction of the biased pixel. Obviously, the particles are pulled into the plasma bubble which is also attributed by a brighter glow of the small plasma bubble if the power increases. At higher power (e.g. ~ *4W*) a very pronounced dome-like glow is formed and the particles are invisible due to the bright plasma. But, suddenly, one after the other particle jumps outwards the bubble to the corners of the pixel where they are collected. Now, the particles are obviously pushed outside the small plasma dome. This "funny" motion of the particles can qualitatively be explained by a dramatic change (reversal) of the electric field in front of the E5 pixel during rf plasma operation with respect to the surrounding field. Vice versa, a decreasing power causes the particles to move back into the bubble. The interaction of dust with a plasma ball by radiofrequency manipulation has also been studied to a certain extent by Annaratone et.al. [59].

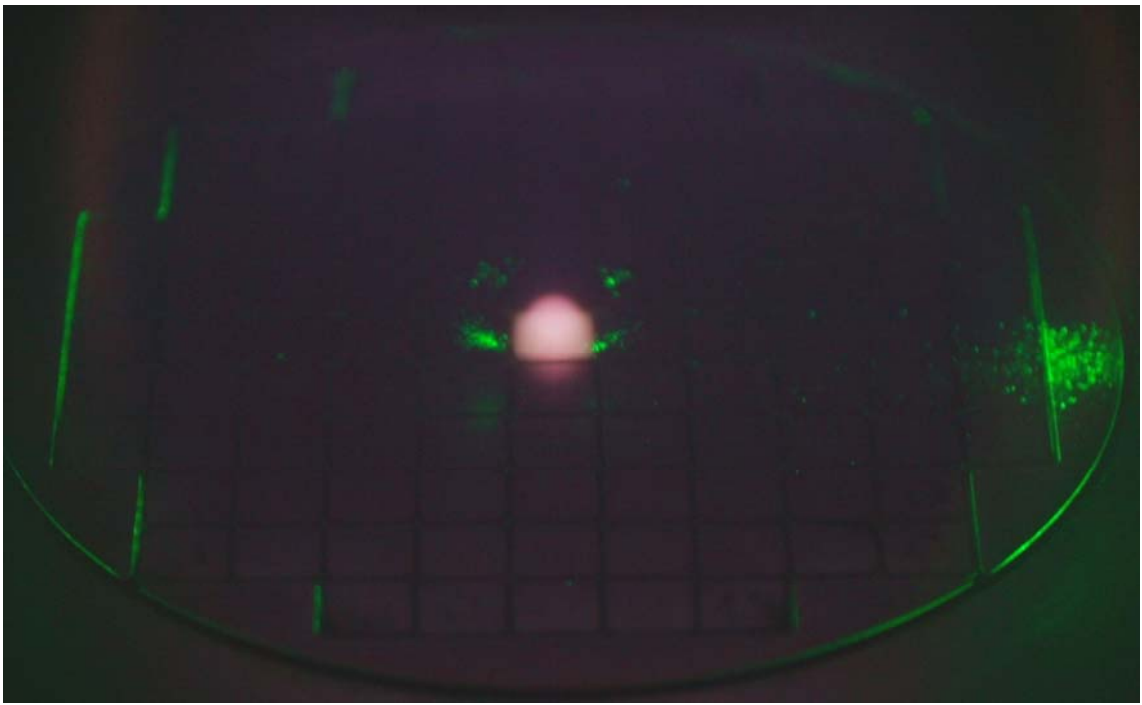

*Fig.16: MF probe particles which are injected into the rf plasma bubble are pushed outwards the bubble and collected in the corners of the pixel.*

The major challenge for a numerical simulation of the particle motion in the vicinity of the light dome is the large range of involved time- and length scales. These range from several *ns* for the rf-frequency over *μs* for the particle charging processes [38, 58] to the actual dynamics of the dust particles which takes place on a time scale of *0.1-100s*. Also to cover all length scales inherent to the system (*μm* dust radius, *mm* sheath thickness, several *cm* vessel dimensions), within one single simulation run seems not practicable due to the numerical effort. But fortunately, this is also not necessary as we can separate the effects on the different scales by using a hierarchical set of models, moderating the numerical requirements. The basic idea behind this approach is to break up the simulation in a part concentrating on the fast dynamics on small length scales, including the effects from larger scales only as fixed input. The results obtained that way then enter the simulation on the coarser time and space grid only in an averaged sense. Depending on the actual time and length scales and the physics involved, different numerical techniques have to be combined.

Note that the goal of our numerical experiment is to test the assumptions on the particle charge in the vicinity of the light dome. Relying on the results from PIC calculations for the plasma characterisation and neglecting the back-action of the probe particles on the plasma, we can compare our results against experimental findings. Using the knowledge about *Q(z)* and *E(z)* obtained in the previous sections to fix the free simulation parameters, we qualitatively reproduce the experimental particle behaviour.

## 3. Summary

In plasma technology it is of great interest to describe the electric field conditions in front of floating or biased surfaces. Therefore we studied the confinement and behaviour of test particles in front of floating surfaces inserted into plasma as well as in front of additionally biased surfaces. The superposition of the electric fields in front of a powered rf-electrode and in front of a perpendicularly orientated floating glass wall results in a force balance for the particles, which causes a wetting behaviour of the glass wall by the particles. The experimental findings can be explained by a simple force model.

Tracking the position and movement of the particles in dependence on the discharge parameters, information on the electric field in front of the electrode has been obtained. These experiments have been performed in front of an adaptive electrode which allows for an efficient confinement and manipulation of the grains. The electric field strength at particle trapping position could be determined to be in the order of several $10^3 V/m$. The experimental results for the obtained electric field in the sheath basically agree with results from independent PIC simulations.

If an additional negative bias potential is suddenly applied to the electrode, the behaviour of a levitated dust particle depends on the magnitude of the bias voltage. At lower voltages the dust particle oscillates around a new equilibrium position. For sufficiently large voltages the particle is accelerated towards the pixel surface by gravity, ion drag and electrostatic forces due to its positive charge, which the particle acquires in absence of electrons.

The response of the plasma to an additional rf-supply at the central pixel has been investigated by a two-dimensional axisymmetric fluid model and two- and three-dimensional PIC simulations. A pronounced enhancement of the plasma becomes obvious from the large increase of the averaged excitation rate and the power density in front of the central pixel. The shape of this enhanced region qualitatively agrees with the visual observations.

Finally, an interesting behaviour of probe particle within the rf plasma bubble has been observed and described qualitatively.

The utilization of charged probe particles in plasma sheath environments is a promising method for the characterization of electric field structures. Suitable experiments in combination with related models provide a powerful tool for the description of plasma structures in the vicinity of electrodes and surfaces. Hence, these experiments and models provide a novel connecting link between sheath diagnostics and plasma wall interaction.

**Acknowledgements**

This work has been supported by the Deutsche Forschungsgemeinschaft under SFB-TR 24. The authors like to thank Andre Melzer, Alexander Mezanov, Ralf Schneider, Gabriele Thieme as well as Josef Blazek for the fruitful discussions and Axel Knuth and Karl-Heinz Schmidt for their technical support.

## 4. References

[1] H. Thomas, G.E. Morfill, V. Demmel V, J. Goree, B. Feuerbacher, D. Möhlmann,
    Phys. Rev. Lett. 73(1994), 652.

[2] A. Bouchoule (Ed.), "Dusty Plasmas", (J.Wiley & Sons, New York, 1999).

[3] A. Melzer, V.A. Schweigert, I.V. Schweigert, A. Homann, S. Peters, A. Piel,
    Phys. Rev. E 54(1996), R46.

[4] New J. Phys. 5(2003), „Focus Issue on Complex (Dusty) Plasmas",
    ed. by G.E. Morfill and H. Kersten – and the papers therein.

[5] V.E. Fortov, O. Petrov, G.E. Morfill, H.M. Thomas et.al.,
    J. Exp. Theor. Phys. 96(2003), 704.

[6] O. Arp, D. Block, A. Piel, A. Melzer, Phys. Rev. Lett. 93(2004), 165004.

[7] S.V. Vladimirov, K. Ostrikov, A.A. Samarian,
    "Physics and Applications of Complex Plasmas", (Imperial College Press, 2005).

[8] A.A. Samarian, B.W. James, Plasma Phys. Control. Fusion 47(2005), B629.

[9] B.M. Annaratone, T. Antonova, H.M. Thomas, G.E. Morfill,
    Phys. Rev. Lett 93(2004), 185001.

[10] J.E. Daugherty, D.B. Graves, J. Vac. Sci. Technol. A11(1993), 1126.

[11] G.H.P.M. Swinkels, H. Kersten, H. Deutsch, G.M.W. Kroesen,
     J. Appl. Phys. 88(2000), 1747.

[12] R. Symes, R.M. Sayer, J. Reid, Phys. Chem. Chem. Phys. 6(2004), 474.

[13] G. Thieme, R. Basner, R. Wiese, H. Kersten, Faraday Discuss. 137(2008), 157.

[14] A.A. Samarian, B.W. James, Phys. Lett. A 287(2001), 125.

[15] C. Zafiu, A. Melzer, A. Piel, Phys. Plasmas 9(2002), 4794.

[16] J.E. Daugherty, R.K. Porteous, D.B. Graves,  J. Appl. Phys., 74(1993), 1617.

[17] J. Goree, Plasma Sourc. Sci. Technol. 3(1994), 400.

[18] L.J. Hou, Y.N. Wang, Acta Phys. Sinica, 52(2003), 434.

[19] L. Boufendi, A. Bouchoule, Plasma Sourc. Sci. Technol. 11(2002), A211.

[20] H. Kersten, G. Thieme, M. Fröhlich, D. Bojic, H.T. Do, M. Quaas, H. Wulff, R. Hippler,
     Pure Appl. Chem. 77(2005), 415.

[21] K.N. Ostrikov, S. Kumar, H. Sugai, Phys. Plasmas 8(2001), 3490.

[22] H. Kersten, H. Deutsch, E. Stoffels, W.W. Stoffels, G.M.W. Kroesen, R. Hippler,
Contrib. Plasma Phys. 41(2001), 598.

[23] H. Kersten, R. Wiese, G. Thieme, M. Fröhlich, A. Kopitov, D. Bojic, F. Scholze,
H. Neumann, M. Quaas, H. Wulff, R. Hippler, New J. Phys. 5(2003), 93.1.

[24] J. Cao, T. Matsoukas, J. Vac. Sci. Technol. B 21(2003), 2011.

[25] E.B. Tomme, B.M. Annaratone, J.E. Allen, Plasma Sourc. Sci. Technol. 9(2000), 87.

[26] H. Kersten, H. Deutsch, G.M.W. Kroesen, Int. J. Mass Spectr. 233(2004), 51.

[27] E. B. Tomme and D. A. Law and B. M. Annaratone and J. E. Allen,
Phys. Rev. Lett. 85(2000), 2518.

[28] S. Ratynskaia, S. Khrapak, A. Zobnin, H. Thomas, M. Kretschmer, A. Usachev,
V. Yaroshenko, R.A. Quinn, G.E. Morfill, O. Petrov, V. Fortov,
Phys. Rev. Lett. 93(2004), 085001.

[29] P.K. Shukla, Phys. Plasmas 8(2001), 1791.

[30] K. Matyash, M. Frohlich, H. Kersten, G. Thieme, R. Schneider, M. Hannemann,
R. Hippler, J. Phys. D: Appl. Phys. 37(2004), 2703.

[31] H. Kersten, E. Stoffels, W.W. Stoffels, M. Otte, K. Csambal, H. Deutsch, R. Hippler,
J. Appl. Phys. 87(2000), 3637.

[32] G. Thieme, V.D. Gorbov, A. Mezanov, R. Wiese, H. Kersten, R. Hippler,
29[th] EPS Plasma Physics, Montreux 2002, ECA 26B(2002) O-4.30.

[33] B.M. Annaratone, M. Glier, T. Stuffler, M. Raif, H.M. Thomas, G.E. Morfill,
New J. Phys. 5(2003), 92.

[34] R. Basner, H. Fehske, H. Kersten, S. Kosse, G. Schubert,
Vakuum in Forschung und Praxis 17(2005), 259.

[35] H. Kersten, R. Wiese, H. Neumann, R. Hippler,
Plasma Phys. Contr. Fusion 48(2006), B105.

[36] A.V. Zobnin, A.D. Usachev, O.F. Petrov, V.E. Fortov, Phys. Plasmas 15(2008), 043705.

[37] M. Tatanova, G. Thieme, R. Basner, M. Hannemann, Y. B. Golubovskii, H. Kersten,
Plasma Sources Sci. Technol. 15(2006), 507.

[38] K. Matyash, R. Schneider, J. Plasma Physics 72(2006), 809.

[39] W.H. Press, B.P. Flannery, S.A. Teukolsky, W.T.Vertterling, "Numerical Recipes",
Cambridge University Press, 1986, Cambridge

[40]  A.A. Samarian, S.V. Vladimirov, Phys. Rev. E 67(2003), 066404.

[41] A. Melzer, S. Nonomura, D. Samsonov, Z. Ma, J. Goree, Phys. Rev. E62(2000) 4162.

[42] A. Barkan, N. d'Angelo, R.L. Merlino, Phys. Rev. Lett. 73(1994) 3093.

[43] C. Zafiu, A. Melzer, A. Piel, Phys. Plasmas 10(2003), 1278.

[44] E. Thomas, B.M. Annaratone, G.E. Morfill et.al., Phys. Rev. E 66(2002), 016405.

[45] C.M. Ticos, A. Dyson, P.W. Smith, Plasma Sourc. Sci. Technol. 13(2004), 395.

[46]  L. Couedel, M. Mikikian, L. Boufendi, Phys. Rev. E74(2006), 026403.

[47]  P. Bryant, J.Phys.D: Appl.Phys. 36(2003), 2859.

[48] S.V. Vladimirov, N.F. Cramer, Phys. Rev. E 62(2000), 2754.

[49] F. Sigeneger, R. Basner, D. Loffhagen, H. Kersten,
     IEEE Trans. Plasma Sci. 36(2008), 1370.

[50] D. Tskhakaya, Lecture Notes in Physics, 739(2008), 168.

[51] http://ptsg.eecs.berkeley.edu

[52] Vahedi, V., C.K. Birdsall, M.A. Lieberman, G. DiPeso, T.D. Rognlien,
     Phys. Fluids B5(1993), 2719.

[53] V. Vahedi , G. DiPeso, J. Comp. Phys. 131(1997), 149.

[54] P. Swarztrauber, SIAM Review 19(1977), 490.

[55] http://www.cisl.ucar.edu/css/software/fishpack

[56] C. Birdsall, IEEE Trans. Plasma Sci. 19(1991), 65.

[57] V. Vahedi, M. Surenda, Comput. Phys. Commun. 87(1995), 179.

[58] F.X. Bronold, H. Fehske, H. Kersten, H. Deutsch, Phys. Rev. Lett. 101(2008), 175002.